\documentclass[prl,superscriptaddress,twocolumn,widetext]{revtex4}
\usepackage{graphicx} 

\def\@biblabel#1{{#1.}}

\def\be{\begin{equation}} \def\ee{\end{equation}}

\newcommand{\ket}[1]{\mbox{$|#1\rangle$}}

\newcommand{\Iphi}{\mbox{$I_{\phi}$}}
\newcommand{\Iuw}{\mbox{$I_{\mu w}$}}
\newcommand{\depth}{\mbox{$\Delta U/\hbar \omega_{p}$}}
\newcommand{\dI}{\mbox{$\delta I$}}
\newcommand{\Idc}{\mbox{$I_{dc}$}}

\begin{document}

\title{Observation of quantum oscillations between a Josephson phase
qubit and a microscopic resonator using fast readout}

\author{K. B. Cooper}
\affiliation{National Institute of Standards and Technology, 325 Broadway, Boulder, CO 80305}
\author{Matthias Steffen}
\affiliation{National Institute of Standards and Technology, 325 Broadway, Boulder, CO 80305}
\affiliation{Center for Bits and Atoms - MIT, Cambridge, MA 02139}
\author{R. McDermott}
\affiliation{National Institute of Standards and Technology, 325 Broadway, Boulder, CO 80305}
\author{R. W. Simmonds}
\affiliation{National Institute of Standards and Technology, 325 Broadway, Boulder, CO 80305}
\author{Seongshik Oh}
\affiliation{National Institute of Standards and Technology, 325 Broadway, Boulder, CO 80305}
\author{D. A. Hite}
\affiliation{National Institute of Standards and Technology, 325 Broadway, Boulder, CO 80305}
\author{D. P. Pappas}
\affiliation{National Institute of Standards and Technology, 325 Broadway, Boulder, CO 80305}
\author{John M. Martinis}
\email{martinis@boulder.nist.gov}
\affiliation{National Institute of Standards and Technology, 325 Broadway, Boulder, CO 80305}



\date{\today}
\begin{abstract}

We have detected coherent quantum oscillations between Josephson phase
qubits and microscopic critical-current fluctuators by implementing a
new state readout technique that is an order of magnitude faster than
previous methods. The period of the oscillations is consistent with the
spectroscopic splittings observed in the qubit's resonant frequency. The
results point to a possible mechanism for decoherence and reduced
measurement fidelity in superconducting qubits and demonstrate the means
to measure two-qubit interactions in the time domain.
 
\end{abstract}

\maketitle

Superconducting circuits based on Josephson tunnel junctions have
attracted renewed attention because of their potential use as quantum
bits (qubits) in a quantum computer\cite{Nielsen00a}. Rapid progress
toward this goal has led to the observation of Rabi oscillations in
charge, flux, phase, and hybrid charge/flux based Josephson
qubits\cite{Martinis02a,Nakamura99a,Chiorescu03a,Vion02a}. Another
milestone toward building a scalable quantum computer is the coherent
coupling of two qubits. While coupled-qubit interactions have been
inferred spectroscopically\cite{Berkley03a,Pashkin03a} and a two-qubit
quantum gate has been implemented\cite{Yamamoto03a}, the direct
detection of correlations of qubit states in the time domain remains to
be demonstrated. One obstacle to observing two-qubit dynamics is that
the single-shot state readout time must be much shorter than the qubit
coherence time ($\sim 10-100$ ns) and the timescale of the coupled-qubit
interaction. Fast readout techniques are also needed for error
correction algorithms\cite{DiVincenzo02}.

Here we report a state measurement of the phase qubit that has high
fidelity and a duration of only $2-4$ ns. Using this new readout
technique, we directly detect time-domain quantum oscillations between
the qubit and the recently discovered spurious resonators associated
with critical-current fluctuators in Josephson tunnel
junctions\cite{Simmonds03a}. These results explicitly illustrate the
mechanism by which critical-current fluctuators decohere phase qubits.
We also present a model that attributes a reduction in measurement
fidelity to the spurious resonators, and we speculate that
qubit-fluctuator coupling contributes to decoherence and loss of
fidelity in the flux and charge/flux qubits as well. In addition to
revealing these new aspects of qubit physics, the few-nanosecond
measurement technique will be valuable for future experiments on coupled
qubits.

\begin{figure}[t]
\begin{center}
\mbox{\includegraphics*[width=3.4in]{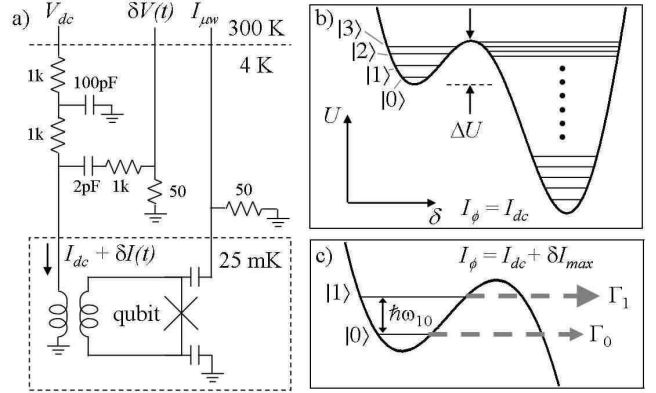}} 
\end{center}
\vspace*{-2.5ex}
\caption{(a) Schematic of the qubit circuitry. For the qubit used in
Fig. 2, the Josephson critical-current and junction capacitance are
$I_{0} \approx 10$ $\mu$A and $C \approx 2$ pF; in Figs. 3 and 4, each
of these values is about 5 times smaller. (b) Potential energy landscape
and quantized energy levels for $\Iphi = \Idc$ prior to the state
measurement. (c) At the peak of $\dI (t)$, the qubit well is much
shallower and state $\ket{1}$ rapidly tunnels to the right hand well.}
\label{fig:Fig1}
\end{figure}

The design and fabrication of the Josephson phase qubits used in this
experiment have been described previously\cite{Simmonds03a,Martinis02a},
and Fig. 1a shows their principal circuitry. The qubit consists of a
superconducting loop containing a single Josephson junction. The qubit
is inductively coupled to a line carrying a flux bias current $\Iphi =
\Idc + \dI (t)$, where $\Idc$ varies slowly and a short pulse $\dI (t)$
is used for the fast readout scheme. The microwave current $\Iuw$
induces Rabi oscillations, and it is capacitively coupled to the qubit
after passing through low-temperature attenuators (not shown). The
dashed box in Fig. 1a surrounds on-chip components kept near 25 mK.
Figure 1b shows the potential energy landscape of the qubit's Josephson
phase. The cubic confinement potential on the left contains several
energy levels, with $\ket{0}$ and $\ket{1}$ representing the qubit
states separated by an energy $\hbar \omega_{10}$. Both $\hbar
\omega_{10}$ and the depth of the left hand well, $\Delta U$, can be
adjusted by varying the bias current $\Iphi$.

Rabi oscillations between states $\ket{0}$ and $\ket{1}$ can be observed
by irradiating the qubit with microwaves at a frequency $\omega/2\pi
\approx \omega_{10}/2\pi \sim 5-10$ GHz and then measuring the qubit's
probability of being in state $\ket{1}$. This probability was previously
measured by applying microwaves at a frequency $\omega_{31}$ for a
duration of 80-100 ns, causing a $1 \rightarrow 3$ transition when
$\ket{1}$ is occupied. Once in state $\ket{3}$, the Josephson phase
rapidly tunnels into the right hand well, and subsequent measurement of
the qubit's flux state with an adjacent dc SQUID (not shown in Fig. 1a)
reveals whether the tunneling occurred. The measurement time of this
method cannot be made significantly shorter than $\sim 80$ ns because of
the need to balance the strength of the $1 \rightarrow 3$ transition
against the tunneling rate of state $\ket{3}$. 

We achieve a faster state measurement of the phase qubit by applying a
short bias current pulse $\dI (t)$ that adiabatically reduces the well
depth $\depth$ so that the state $\ket{1}$ lies very near the top of the
well when the current pulse is at its maximum $\dI_{max}$ (see Fig. 1c).
The value of $\dI_{max}$ is chosen so that the tunneling rate
$\Gamma_{1}$ of state $\ket{1}$ at $\dI_{max}$ is high enough that
$\ket{1}$ will almost certainly tunnel over the duration of $\dI (t)$.
Also, because $\Gamma_{1}$ is at least two orders of magnitude larger
than the tunneling rate $\Gamma_{0}$ of $\ket{0}$, a single current
pulse yields a reliable measurement of the probability that $\ket{1}$ is
occupied. Calculations suggest that the ratio of tunneling rates for
shallow wells is $\alpha = \Gamma_{1}/\Gamma_{0} \approx 150$, and that
the corresponding maximum measurement fidelity is $\eta \approx
\alpha^{-1/\alpha} - \alpha^{-1} \approx 0.96$. Here $\eta$ is defined
as the difference of the tunneling probability when the qubit is in
state $\ket{1}$ versus state $\ket{0}$.

\begin{figure}[t]
\begin{center}
\mbox{\includegraphics*[width=3.4in]{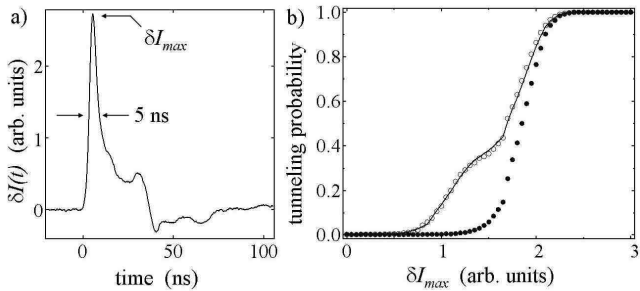}} 
\end{center}
\vspace*{-2.5ex}
\caption{(a) Room temperature measurement of the fast current pulse. (b)
Tunneling probability versus $\dI_{max}$ with the qubit in state
$\ket{0}$ (solid circles) and in an equal mixture of states $\ket{1}$
and $\ket{0}$ (open circles). Fit to data is shown by the solid line.
The plateau, being less than 0.5, corresponds to a maximum measurement
fidelity of 0.63.}
\label{fig:Fig2}
\end{figure}

Room temperature measurements reveal that $\dI (t)$ has a width of about
5 ns, as shown in Fig. 2a. This is sufficiently slow to maintain
adiabaticity with respect to the subnanosecond time scales associated
with intrawell transitions. The actual measurement time will be somewhat
shorter than the full width of $\dI (t)$ because the tunneling rate
$\Gamma_{1}$ is \emph{exponentially} sensitive to the total bias current
$\Iphi$. Therefore, the qubit will be far more likely to tunnel near the
peak of $\dI (t)$ rather than its flanks, including the long
trailing-edge of $\dI (t)$ arising from impedance mismatches in the
current bias line. We estimate the effective measurement duration to be
only $2-4$ ns. This is more than an order of magnitude shorter than the
previously used microwave measurement technique as well as the readout
methods used in most other superconducting qubits\cite{DelftNote}.

\begin{figure}
\begin{center}
\mbox{\includegraphics*[width=3.4in]{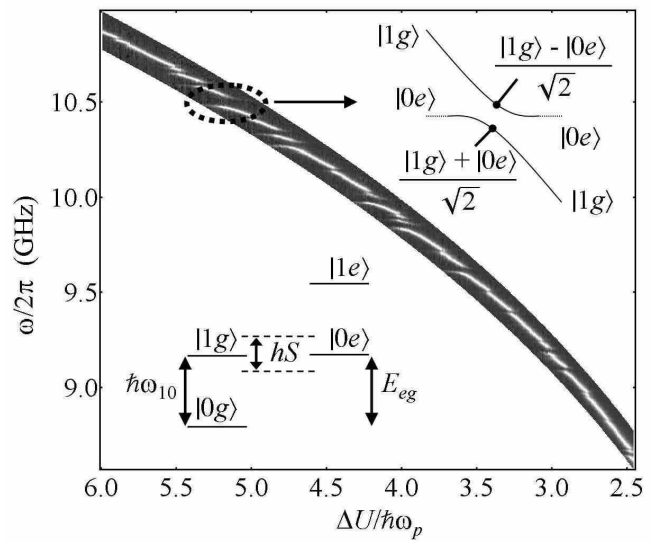}} 
\end{center}
\vspace*{-2.5ex}
\caption{Spectroscopy of $\omega_{10}$ obtained using the current-pulse
measurement method, as a function of well depth $\depth$. For each value
of $\depth$, the grayscale intensity is the normalized tunneling
probability, with an original peak height of $0.1-0.3$. Insets: A given
splitting in the spectroscopy of magnitude $S$ comes from a
critical-current fluctuator coupled to the qubit with strength $hS/2$.
On resonance, the qubit-fluctuator eigenstates are linear combinations
of the states $\ket{1g}$ and $\ket{0e}$, where $\ket{g}$ and $\ket{e}$
are the two states of the fluctuator.}
\label{fig:Fig3}
\end{figure}

The data in Fig. 2b demonstrate the effectiveness of the fast
measurement scheme. The solid circles correspond to the tunneling
probability of the qubit as a function of the pulse height $\dI_{max}$
when no radiation at $\omega_{10}$ is applied, i.e., when the qubit is
in state $\ket{0}$. The data were obtained at an initial well depth of
$\Delta U = 4.5\hbar\omega_{p}$, where $\omega_{p} \approx
\omega_{10}/0.9$ is the classical plasma frequency of the Josephson
junction. The open circles in Fig. 2 are the measured tunneling
probabilities as a function of $\dI_{max}$ after a microwave drive at
$\omega_{10}$ saturates the populations of $\ket{0}$ and $\ket{1}$
approximately equally. To produce a $\approx 50/50$ mixture of $\ket{0}$
and $\ket{1}$, microwaves were applied for 500 ns, much longer than the
qubit's $T_{1}$ time, and their power was high enough that the Rabi
oscillation period of about 10 ns is shorter than $T_{1}$. The plateau
in the tunneling probability data occurs around the values of
$\dI_{max}$ where state $\ket{1}$ has a high tunneling rate while state
$\ket{0}$ remains mostly confined in its potential well. For equal
populations of $\ket{0}$ and $\ket{1}$, the plateau should level out
near 0.50 for the predicted measurement fidelity of $\eta = 0.96$.
Instead, the measured tunneling probability plateaus around 0.35,
suggesting a lower, but still good, fidelity. Indeed, fitting the
tunneling data to a simple model with a constant tunneling ratio
$\alpha$ yields a maximum fidelity of $\eta = 0.63$ (solid line in Fig.
2). One possible reason for this lower fidelity will be discussed below.

The new state readout scheme is capable of measuring the spectroscopy of
the $0 \rightarrow 1$ transition for a broad range of well depths, as
shown in Fig. 3. The data were obtained from a qubit with a slightly
lower fidelity ($\eta \approx 0.5$) than that of Fig. 2b, but both
exhibit the same essential behavior. Similar spectroscopic data were
shown in ref. \cite{Simmonds03a}, and the grayscale is proportional to
the occupation probability of state $\ket{1}$ after a long, low power
microwave drive is applied. However, here the fast-pulse measurement
method is used with $\delta I_{max}$ adjusted to optimize the signal at
each flux bias point. The new measurement method is capable of probing a
much broader range of $\depth$ than the old method, where the slow
tunneling rate of $\ket{3}$ limits the accessible well depths to less
than about $\depth = 3.5$. Also, Fig. 3 shows a series of gaps in the
$\omega_{10}$ resonance signal, and these splittings likely reflect
avoided level crossings arising from an interaction of the qubit with
individual critical-current fluctuators at microwave
frequencies\cite{Simmonds03a}. Treating a single fluctuator as two-level
quantum systems and labeling its ground and excited states as $\ket{g}$
and $\ket{e}$, a coupling of strength $hS/2$ will split the
direct-product states $\ket{1g}$ and $\ket{0e}$ by $hS$ when the qubit
energy $\hbar \omega_{10}$ is tuned to the fluctuator energy $E_{eg}$
(see insets to Fig. 3). Splittings as large as $S \approx 70$ MHz are
visible in Fig. 3.

\begin{figure}[t]
\begin{center}
\mbox{\includegraphics*[width=3.4in]{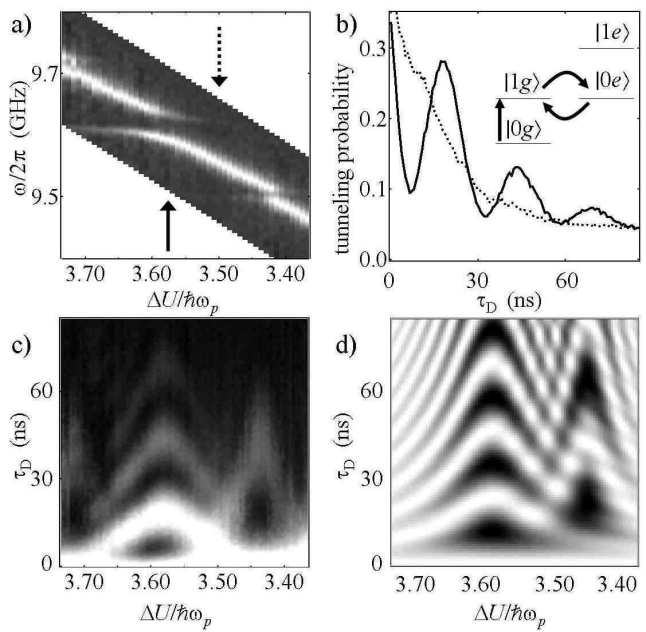}} 
\end{center}
\vspace*{-2.5ex}
\caption{(a) Detail of the qubit spectroscopy near $\depth = 3.55$,
showing splittings of strengths $S \approx$ 44 MHz and 24 MHz. (b)
Tunneling probability versus measurement delay time $\tau_{D}$ after
application of $\pi$-pulse. Solid (dashed) line is taken at a well depth
of solid (dashed) arrow in (a), corresponding to a resonant
(off-resonant) bias. Inset illustrates how the qubit probability
amplitude first moves to state $\ket{1g}$ and then oscillates between
$\ket{1g}$ and $\ket{0e}$. (c) and (d) Tunneling probability (gray
scale) versus well depth and $\tau_{D}$ for experimental data (c) and
numerical simulation (d). The peak oscillation periods are observed to
correspond to the spectroscopic splittings.}
\label{fig:Fig4}
\end{figure}

Simmonds \emph{et al.} have already shown that the qubit's Rabi
oscillations have reduced coherence when $\omega_{10}$ is tuned near a
splitting in the spectroscopy\cite{Simmonds03a}. To better understand
the spurious resonators' effect on the qubit, it is helpful to examine
the \emph{dynamics} of the qubit-fluctuator interaction directly, and
the few-nanosecond readout method allows us to accomplish this. Figure
4a shows a section of the spectroscopy of Fig. 3 around $\depth = 3.6$,
where a strong, well-isolated splitting occurs at $\omega_{10}/2\pi =
9.62$ GHz with a magnitude of $S \approx 44$ MHz. A smaller splitting of
magnitude $S \approx 24$ MHz is visible at a slightly shallower well
depth. Figure 4b shows the time-domain response of the qubit to an 8 ns
$\pi$-pulse for the qubit tuned to the center of (solid) and away from
(dashed) the 44 MHz splitting in Fig. 4a. Following the $\pi$-pulse, the
fast measurement probe is applied after a delay of $\tau_{D}$ to measure
how the occupation probability of $\ket{1}$ changes with time. For a
well depth $\depth = 3.50$, the dashed curve in Fig. 4b exhibits an
exponential decay with a time constant that is roughly $T_{1} \approx
25$ ns\cite{T1Note}. In contrast, the solid curve in Fig. 4b shows that
when the qubit is tuned to a large splitting, at $\depth = 3.58$, a
striking beating in the tunneling probability is superimposed on the
$T_{1}$ decay curve. Although this beating represents a probability
oscillation between $\ket{1}$ and $\ket{0}$, it is \emph{not} a Rabi
oscillation because there is no microwave driving power at
$\omega_{10}$. Instead, the beating period of 24 ns is very close to the
inverse of the measured splitting size $S^{-1} = 23$ ns, which is
expected from the model of the qubit coupled to a critical-current
fluctuator with a strength $S/2$. As shown in the inset to Fig. 4b,
after the qubit is promoted to state $\ket{1g}$ by the $\pi$-pulse, the
qubit-fluctuator interaction will cause an oscillation between
$\ket{1g}$ and $\ket{0e}$ at a frequency $S$ as energy is transferred
back and forth between the qubit and the fluctuator. This constitutes
compelling evidence for coherent quantum oscillations between the
mesoscopic qubit and a single microscopic fluctuator. 

A further test of this model is to track the time-domain response of the
qubit over a narrow range of bias currents around the fluctuator's
resonant frequency. As the qubit bias is moved away from resonance, we
expect the beating frequency to increase as the states $\ket{1g}$ and
$\ket{0e}$ become nondegenerate. This expectation is confirmed in Fig.
4c, which shows that the longest beating periods line up well with the
major and minor splittings visible in Fig. 4a. By numerically modeling
the qubit-fluctuator interaction based on the location and sizes of the
splittings in Fig. 4a, we obtain the results in Fig. 4d (where
dissipation is ignored). The simulation matches the experimental results
remarkably well, except for the exponential signal decay in Fig. 4c, as
expected. However, note that our simulations and experimental results do
not rule out the possibility that the resonator might have other excited
levels out of resonance with the qubit. 

We emphasize that these results could not have been obtained using the
previous microwave measurement method because the signals would be
averaged out over the $\sim$100 ns readout time. Probability beatings on
longer time scales coming from more weakly coupled resonators could not
be resolved either because of the qubit's short $T_{1}$ times. The same
limitations will exist if attempts are made to directly couple two
qubits. Therefore, the new fast measurement presented here and the
demonstration of dynamical coupling between the qubit and a
critical-current fluctuator suggest that we now have the tools to
successfully measure the coupling between two Josephson phase qubits.

The data of Fig. 4 demonstrate explicitly how a single, large fluctuator
will absorb the occupation probability of state $\ket{1}$.
Interestingly, once the fluctuator absorbs the qubit energy after $\sim
10$ ns, it does not immediately decay incoherently from $\ket{0e}$ to
$\ket{0g}$. In fact, the envelope of decay of the beating signal in Fig.
4b is about one to two times the $T_{1}$ of the qubit away from a large
resonator. This means that the decay time of a strong critical-current
fluctuator is at least as long as the qubit's $T_{1}$ time. We thus
speculate that spin-echo techniques might be able to refocus some of the
signal loss due to the spurious resonators\cite{Linden99d}. Another
feature of the qubit-fluctuator interaction still to be explored is the
effect of many small fluctuators not resolved in the spectroscopy data.
Analyses of the resonator distributions could reveal how strongly such
an ensemble of coupled critical-current fluctuators would affect the
qubit and whether this might be a factor in the short $T_{1}$ observed.

Another consequence of the dynamic qubit-resonator interaction is
reduced fidelity of the fast-pulse measurement itself. As $\dI (t)$
increases during a measurement, $\omega_{10}$ decreases and the qubit
moves in and out of resonance with many spurious resonators before any
tunneling occurs. If the qubit is initialized in state $\ket{1}$, then
each resonator absorbs a small amount of the $\ket{1}$ probability
amplitude during the measurement pulse, leaving the qubit with some
amplitude in state $\ket{0}$. The probability of remaining in state
$\ket{1}$ after sweeping through a single fluctuator of strength $hS/2$
can be estimated from the Zener-Landau tunneling formula $P(S) =
\rm{exp}$$(-\pi^2S^2/ \dot f_{10})$, where $\dot f_{10} = \dot
\omega_{10}/2 \pi$ is the rate of change of the qubit frequency during
the sweep\cite{Zener}. 

Accounting for the effect of a collection of $N_{S_{i}}$ resonators of
splitting size $S_{i}$, the total measurement fidelity becomes $\eta
\approx \prod_{i} P(S_{i})^{N_{S_{i}}}$. For the qubit used in Fig. 2,
spectroscopic measurements indicate that the rms splitting size of the
45 visible splittings is $S_{rms} \approx 30$ MHz. Assuming that $\delta
I(t)$ results in a frequency sweep rate of $\dot f_{10} \approx 1$
GHz/ns, we find that the measurement fidelity would be reduced from
$\eta \approx 1$ to $\eta \approx 0.7$. The actual fidelity of the qubit
of Fig. 3 is $\eta = 0.63$, and therefore the qubit-fluctuator
interaction is likely a prominent source of fidelity loss in the
fast-pulse measurement method. Somewhat counter-intuitively, the
Landau-Zener model predicts that the fidelity should become \emph{worse}
as the measurement duration becomes \emph{longer}. Preliminary
experiments involving slower pulses of $\delta I(t)$ are consistent with
this prediction, but so far it has been difficult to separate the effect
of fidelity loss from the signal loss due to short $T_{1}$ times.
Nonetheless, this effect may have consequences for other superconducting
qubit systems because similar current pulse schemes for state
measurement and manipulation have been employed in the flux and the
charge/flux qubits\cite{Chiorescu03a,Vion02a}. Both qubits also exhibit
significantly reduced fidelities of $\eta \sim 0.6$. However, in the
case of the charge qubit a probability plateau analogous to that of Fig.
2b is absent, and the fidelity is estimated indirectly from the
amplitude of Rabi oscillations\cite{PlateauNote}. Whether these lowered
fidelities can be also attributed to microscopic fluctuators remains to
be investigated.

In conclusion, we have implemented a state measurement technique for the
Josephson phase qubit that is an order of magnitude faster than the
microwave measurement method. With a temporal resolution of less than 5
ns, the fast-pulse method reveals coherent quantum oscillations between
the qubit and a microscopic resonator embedded within the qubit circuit.
The dynamics of the qubit-resonator interaction illustrate one mechanism
by which the coherence of a superconducting qubit is lost to its
environment. The size and number of the resonators also suggest that
they are relevant to fidelity loss in pulse measurements, and we predict
that the fidelity should increase as the measurement duration decreases.
These results underscore the importance of understanding the details of
Josephson junction physics in order to explain the quantum behavior of
superconducting qubits. They also prove that all the tools are available
for a time-domain demonstration of the coupling of two superconducting
phase qubits.

\begin{acknowledgments}
This work was supported in part by NSA under contract MOD709001.
\end{acknowledgments}


\begin{thebibliography}{1}
\bibitem{T1Note}
The expected $T_{1}$ times are on the order of microseconds, and the cause of the fast energy relaxation is under investigation.
\end{thebibliography}
\end{document}